\documentclass[aps,onecolumn,preprint,superscriptaddress
]{revtex4-1}
\pdfoutput=1
\usepackage{amsmath}
\usepackage{amssymb}
\usepackage{subfig}
\usepackage{graphicx}
\usepackage{float}
\usepackage{tabularx}
\usepackage{hyperref}
\usepackage{physics}
\usepackage{bm}

\graphicspath{{"./figures/"}{"./supplemental/figures/"}}

\begin{document}

\title{ \textbf{On the question of second sound in germanium: A theoretical viewpoint}}
\author{ {{Samuel Huberman}}}
\email[]{samuel.huberman@mcgill.ca}
\affiliation{Department of Chemical Engineering, McGill University, 
Montreal, Quebec H3A 0C5, Canada}

\author{ {{Chuang Zhang}}}
\affiliation{Department of Mechanics and Aerospace Engineering, Southern University of Science and Technology, Shenzhen 518055, China}

\author{ {{Jamal Abou Haibeh}}}
\affiliation{Department of Chemical and Biological Engineering, University of Ottawa, 161 Louis
Pasteur, Ottawa, Ontario, K1N 6N5, Canada}
\begin{abstract}
We revisit the recent work from Beardo et al.~\cite{beardo2021observation} wherein the observation of second sound in germanium is claimed. We review the requirements imposed on the collision operator (or equivalently, the full scattering matrix) of the linearized phonon Boltzmann transport equation (LBTE) for the observation of driftless second sound as established by Hardy. By performing an eigendecomposition of the full scattering matrix, we show that the requirement that the smallest nonzero eigenvalue must be associated with an odd eigenvector is not satisfied. Furthermore, direct solutions to the LBTE for a frequency modulated heat source do not reveal the presence of second sound. Finally, numerical solutions to the BTE under the relaxation time approximation (RTA) in the 1D frequency-domain thermoreflectance (1D-FDTR) experimental geometry demonstrate that phase lag alone is not a suitable experimental observable for inferring second sound. We conclude by discussing the need for a second sound `smoking gun'.

\end{abstract}
\maketitle

\section{Introduction}

Second sound is a manifestation of phonon hydrodynamics. Peshkov theoretically proposed that phonon hydrodynamics can occur in crystal solids~\cite{peshkov1944second}. This hypothesis was subsequently confirmed by a handful of experiments reporting the observation of second sound in a few materials at low temperatures~\cite{ackerman1966second,jackson1970second,narayanamurti1972observation}. Further theoretical worked by Enz categorized the distinct speeds of second sound as `drifting' and `driftless' second sound~\cite{enz1966microscopic}. 

Hardy subsequently showed that these two distinct manifestations of wave-like temperature responses in crystals can be obtained from different approximate solutions to the LBTE, so long as certain requirements imposed on the eigenvalues of  collision operator are satisfied~\cite{hardy1970phonon}. Hardy further noted that `other types' of second sound may exist so long as the eigenvalues of the collision operator yield a slow decay of energy flux relative to the heating frequency.

In Ref. ~\cite{beardo2021observation}, the frequency domain thermoreflectance (FDTR) technique is used to study thermal transport in a bulk germanium crystal. In this technique, an intensity modulated optical pump beam acts as frequency modulated thermal source and a probe beam encodes changes in reflectivity (which is assumed to be directly proportional to changes in temperature) at the surface of the sample. The reported experimental observable is the phase lag between the heat source and the temperature response at the surface. Over a temperature range from 7 K to 300 K, the experimental phase lag is found deviate to from the diffusion equation at high frequencies. To interpret the data, a macroscopic hyperbolic heat equation (HHE) is invoked for which the solutions for temperature are damped waves:

\begin{widetext}
\begin{equation}
\tau\frac{\partial^2 T}{\partial t^2} + \frac{\partial T}{\partial t} -\alpha \nabla^2 T = \frac{1}{\rho C_p}\left(S(r,t) + \tau\frac{\partial S (r,t)}{\partial t}\right),
\end{equation}
\end{widetext}

where $\tau$ denotes a macroscopic relaxation time, $T$ is the temperature, $\alpha$ is the thermal diffusivity, $\rho$ is density, C$_p$ is specific heat and $S(r,t)$ is an external heat source. Reasonable fits between this equation and the experimental data is considered evidence for `driftless' or `other types' of second sound (due to the weak normal scattering rate relative to the Umklapp and isotopic scattering rates in germanium, drifting second sound will not appear, see Supplemental Material). However, a microscopic picture of the experimental data is still lacking. Using the BTE, we take a step in addressing this missing piece in this paper, which is organized as follows. In Sec~\ref{sec:2}, an eigenvalue analysis of the collision operator is performed. In Sec.~\ref{sec:3}, direct solutions to the LBTE are obtained and compared with the diffusion, HHE and ballistic solutions. Finally, in Sec.~\ref{sec:4}, results of numerical solutions to the BTE under the RTA in the 1D-FDTR geometry are presented and compared with the diffusion and HHE predictions.

\section{Eigendecomposition of collision operator}\label{sec:2}

Past studies of driftless second sound were limited to a purely mathematical realm; Hardy stopped short of making predictions because of the inability to obtain the matrix elements of the collision operator. As did others, to quote Beck et al. `In view of the analysis performed in~\cite{meier1973displacement}, we conclude that this driftless second sound occurs if the energy current density is an approximately conserved variable, which does not couple to the momentum density. In reality, however, the energy flux can be decomposed in a component which is attenuated by normal processes and one which decays due to U-processes. The latter part is expected to be strongly coupled to the momentum density, and this coupling then gives rise to the usual concept of drifting second sound. A formal analysis of the eigenvalues and eigenvectors of the collision operator performed by Hardy shows that mathematically it is possible that the coupling is negligible but the necessary conditions on the eigenvalue spectrum cannot be interpreted in physical terms. A definite conclusion on the feasibility of driftless second sound should, therefore, be subject to explicit results on the eigenvalues of the collision operator.' ~\cite{beck1974phonon} In recent years, with the success of ~\textit{ab initio} methods~\cite{fugallo2013ab}, such `explicit results' have become obtainable.

So, in the spirit of the above statement, we investigate the eigenvalue spectrum of the collision operator of germanium to ascertain the feasibility of observing driftless second sound. The eigenvalue problem is expressed as

\begin{figure}
\centering     
\subfloat[]{\label{fig:1a}\includegraphics[width=0.45\textwidth]{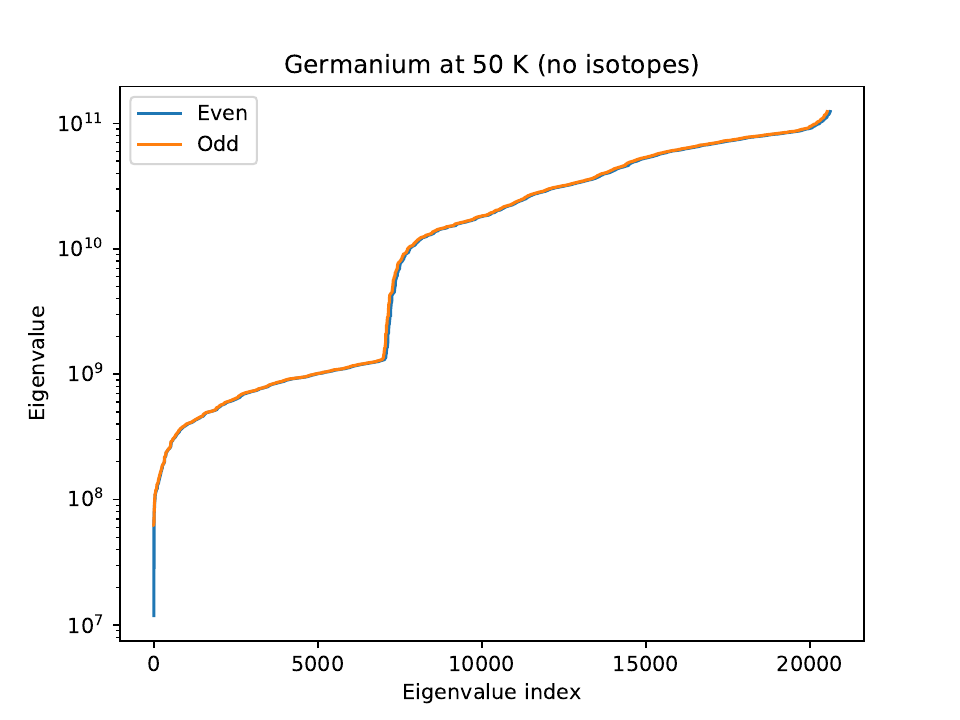}}
\subfloat[]{\label{fig:1b}\includegraphics[width=0.45\textwidth]{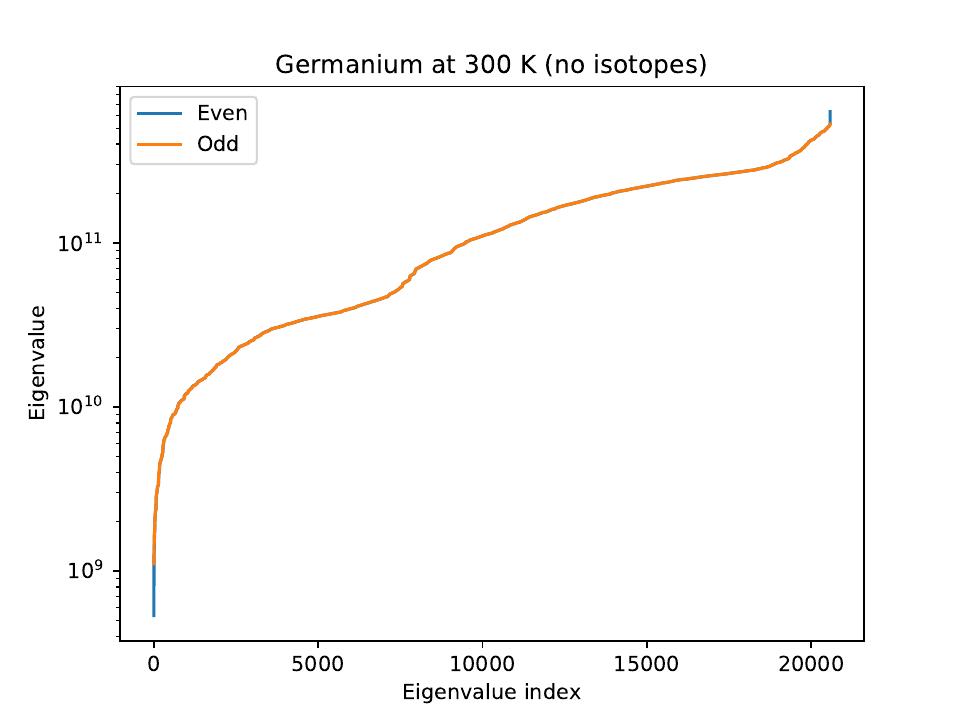}}
\caption{Even and odd eigenvalues of the germanium collision operator.}\label{fig:1}
\end{figure}

\begin{widetext}
\begin{equation}
\sum_{n'}\tilde{\Omega}_{nn'} \theta^{\alpha}_{n'} = \frac{1}{\tau_{\alpha}}\theta^{\alpha}_{n'},
\end{equation}
\end{widetext}

where $\tilde{\Omega}_{nn'}$ is the symmetrized collision operator, $\theta^{\alpha}_{n'}$ is $n'$ component of the $\alpha$ eigenvector with the corresponding eigenvalue $\frac{1}{\tau_{\alpha}}$. Hardy formulated the requirements for driftless second sound in terms of the eigenvalues of the collision matrix: the smallest nonzero eigenvalue must be associated with an odd eigenvector~\cite{hardy1970phonon} for it is the odd eigenvectors that contribute to heat flux while the even eigenvectors contribute to energy change. Here, even and odd is defined by the sign of the eigenvector component upon changing the sign of the phonon wavevector $\bm{q}$:  $\theta^{\alpha}_{n} = \pm\theta^{\alpha}_{-n}$ where $-n = (-\bm{q},b)$ where $b$ denotes the phonon branch. Following the procedure outlined in Ref.~\cite{cepellotti2016thermal} using~\textit{ab initio} properties for germanium~\cite{huberman2017unifying}, we have obtained the eigenvalue spectrum for the collision matrix of pure germanium at $T$ = 50 K and 300 K and find that this condition is~\textit{not} satisfied: the smallest odd eigenvalue is larger than the smallest even eigenvalue (see Figure~\ref{fig:1}). Similar results were previously obtained for silicon (see Figure 2c in Ref.~\cite{simoncelli2020generalization}).

\section{LBTE solutions}\label{sec:3}

In previous studies, several different ansatzs for the phonon distribution functions have been used to anticipate these `other types' of second sound. Cepellotti et al. assume plane wave forms for the non-equilibrium distribution~\cite{cepellotti2017transport}. Sendra et al. assume a form where the heat flux and its first derivatives in time and space are considered independent variables~\cite{sendra2021derivation}. Recently, Shang et al. used a first order correction to the drifted Bose-Einstein distribution under the Callaway approximation to the collision operator to obtain a dispersion relation for second sound~\cite{shang2022unified}. Here, we do not impose any ansatz on the phonon distribution function and in doing so retain the ability to capture Hardy's three types of second sound without assuming \textit{a priori} their occurrence. Assuming the temperature rise due to the heating is small compared to the background temperature such that a local temperature exists, we directly solve the linearized BTE as previously presented in Ref.~\cite{chiloyan2021green}.

\begin{equation}\label{Eq:3}
\frac{\partial g_n}{\partial t} + \bm{\vec{v}}_n \cdot \nabla g_n  = \sum_{j} \Omega_{n,n'}\frac{\omega_n}{\omega_{n'}}(g_{n'}^0-g_{n'})+Q_n,
\end{equation}

where $g_n$ is the deviational phonon energy density for mode $n$, $\bm{\vec{v}}_n$ is the phonon group velocity, $\Omega_{n,n'}$ is the unsymmetrized collision operator, $\omega_n$ is the phonon frequency, $g_{n'}^0$ is the deviational equilibrium phonon energy density and $Q_n$ is the spectral heat source. Using the Green's functions solutions to Eq.~\ref{Eq:3}, we calculate the temperature response for the case where $Q= \bar{Q}\delta(x)e^{iwt}$: a frequency-dependent point heat source. One can obtain experimentally-relevant heat sources, such as a Gaussian hot spot, via linear combinations of the solutions presented here.

\begin{figure}
\centering     
\subfloat[]{\label{fig:1a}\includegraphics[width=0.5\textwidth]{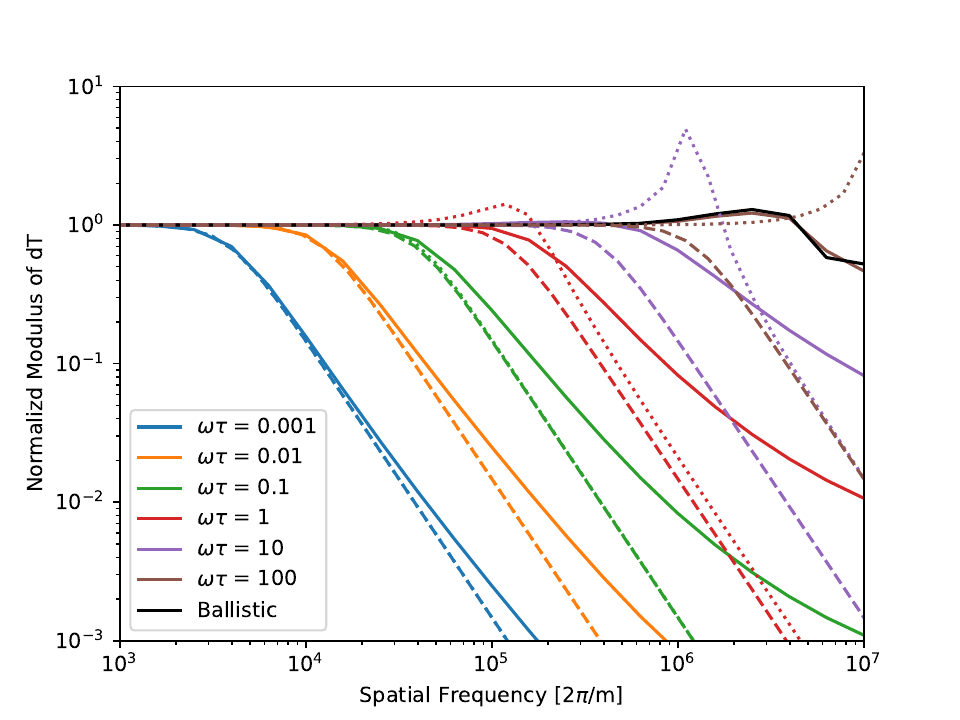}}
\subfloat[]{\label{fig:1b}\includegraphics[width=0.5\textwidth]{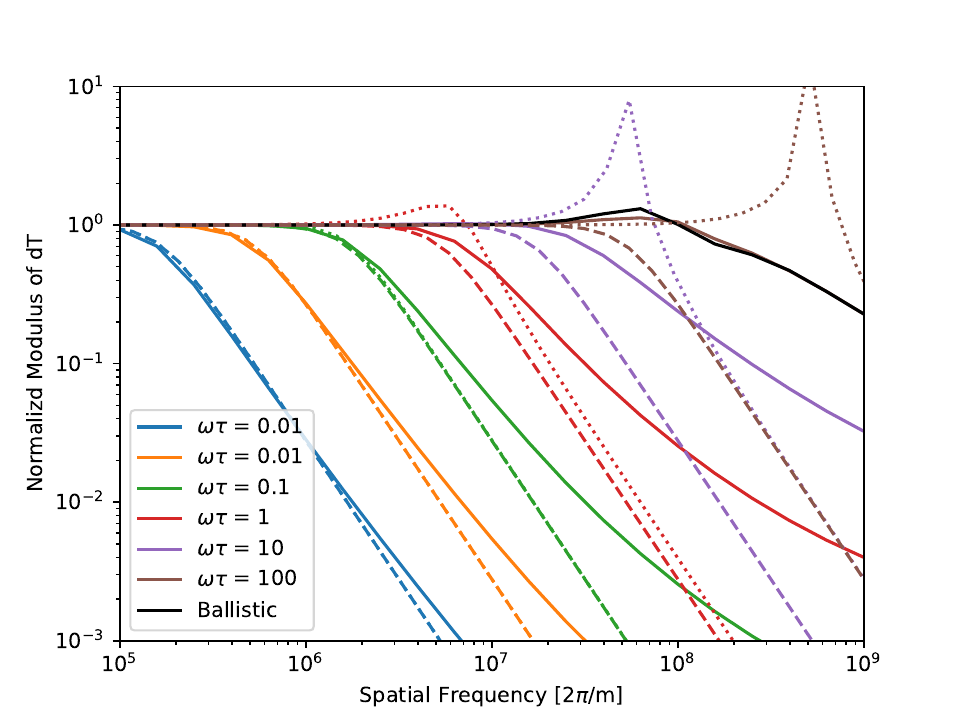}}
\caption{Temperature response for $Q= \bar{Q}\delta(x)e^{iwt}$ in germanium at (a) $T$ = 50 K and (b) $T$ = 300 K. Dashed, dotted and solid lines correspond to the diffusion, HHE and LBTE solutions respectively. The black line is the ballistic solution for $\omega \tau = 100$.}\label{fig:2}
\end{figure}

The spatial-frequency domain temperature responses for germanium at $T$ = 50 K and 300 K are presented in Figure~\ref{fig:2}. At low heating frequencies ($\omega \tau < 0.001 $), the diffusive limit is recovered by the HHE and LBTE solutions. At $\omega \tau \approx$ 0.01 to 0.1 , the LBTE begins to significantly deviate from the diffusion and HHE solutions. At $\omega \tau \approx$ 1 to 10, the HHE yields a resonant feature, while the LBTE does not. At $\omega \tau >$ 100, the ballistic regime is approached by the LBTE but not by the HHE. The absence of a resonance in the LBTE solutions for $\omega \tau >$ 1 indicate that temperature waves are negligible in germanium. Additionally, we have calculated the temperature response for $Q= \bar{Q}\delta(t)e^{iqx}$ (1D-TTG), which do not yield evidence of driftless or any `other type' of second sound (see Supplemental Material).

When the same calculations are performed with graphite, an unambiguous contrast is found. The LBTE spatial frequency domain temperatures responses for graphite at $T$ = 100 K are presented in Figure~\ref{fig:3}. At low heating frequencies (10 MHz), the temperature response is diffusive-like. At intermediate heating frequencies (100 - 500 MHz), a resonance is observed. This resonance can be associated with the previously reported observation of second sound in graphite~\cite{huberman2019observation}. At high heating frequencies (5 - 25 GHz), the intensity of this resonance is reduced. In other words, in graphite, a clear transition from the diffusive to hydrodynamic to ballistic regimes is observed. 

\begin{figure}
\includegraphics[width=0.5\textwidth]{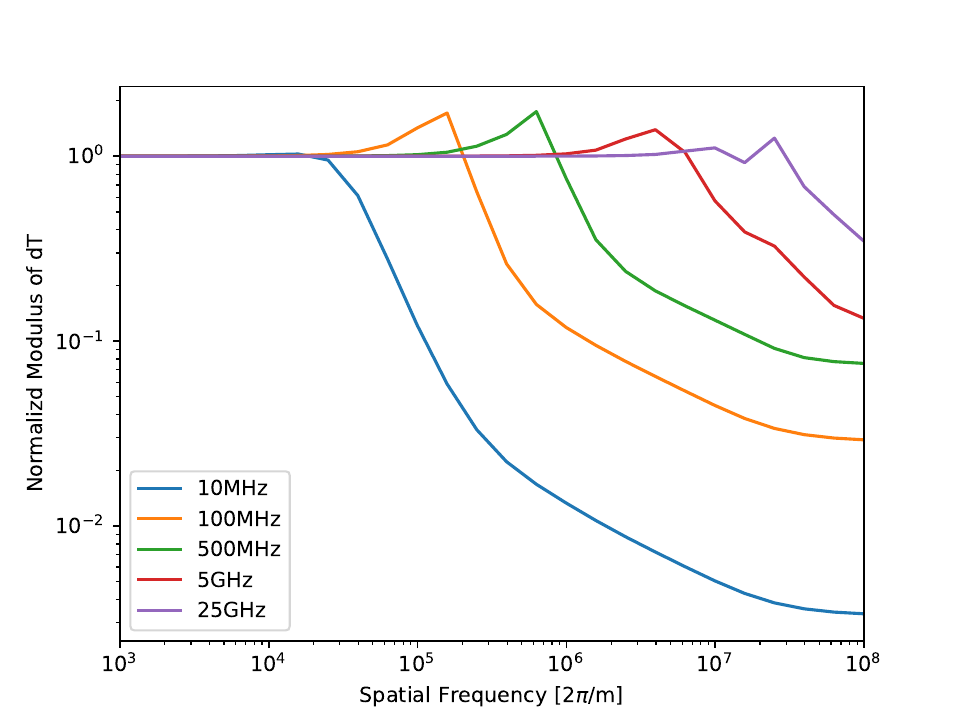}
\caption{LBTE Temperature response in graphite $T$ = 100 K for $Q= \bar{Q}\delta(x)e^{iwt}$.}
\end{figure}\label{fig:3}

It can be argued that at very high heating frequencies ($\omega \tau >> 1$), regardless of the specifics of the collision operator, that wave-like transport can exist. This is the temporal equivalent to the ballistic limit. Wave-like ballistic transport has been observed in the original second sound heat pulse experiments in NaF~\cite{jackson1970second}. Indeed, we do observe resonances in the ballistic limit of graphite (Figure~\ref{fig:3} at $f$ = 25 GHz as well as the Supplemental Material and Ref.~\cite{ding2022observation}). These ballistic temperature waves can be distinguished from second sound oscillations through inspection by making a suitable correspondence to the first sound speeds of the material.

\section{Direct simulations of the FDTR geometry}\label{sec:4}

The solutions of the full scattering matrix LBTE are presently restricted to unbound geometries or specularly reflecting boundaries~\cite{hua2014analytical}. To simulate a geometry representative of the FDTR experiment with a diffuse boundary, we turn to the Discrete Unified Gas Kinetic Scheme (DUGKS) that is capable of solving the BTE under the RTA and Callaway approximations~\cite{zhang2017unified} with \textit{ab initio} inputs (see Supplemental Material).

\begin{figure}
\centering     
\subfloat[]{\label{fig:a}\includegraphics[width=0.45\textwidth]{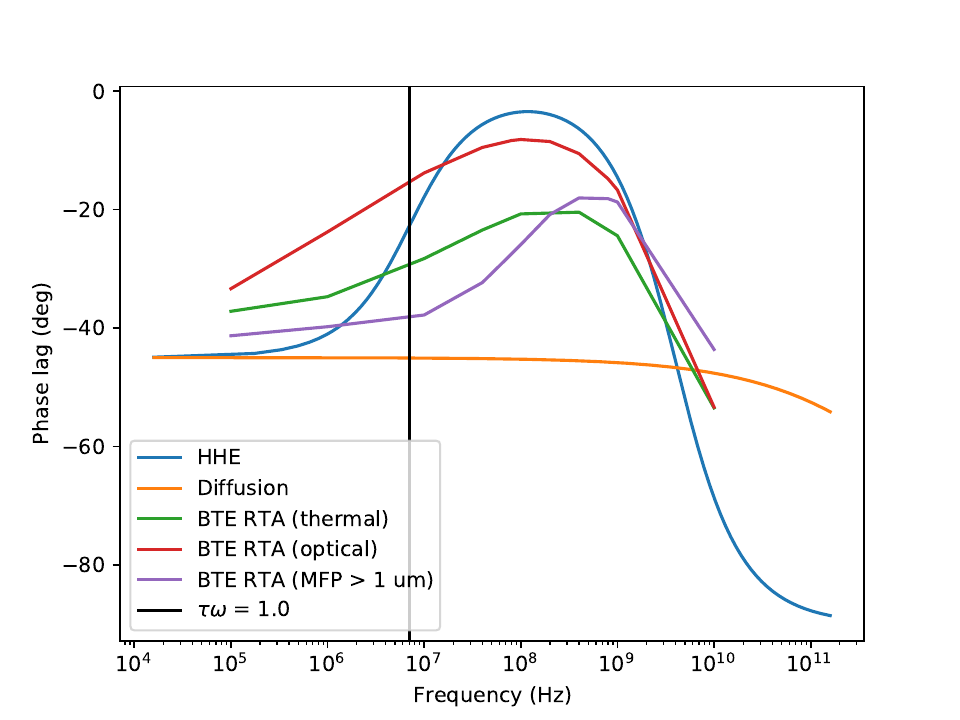}}
\subfloat[]{\label{fig:b}\includegraphics[width=0.45\textwidth]{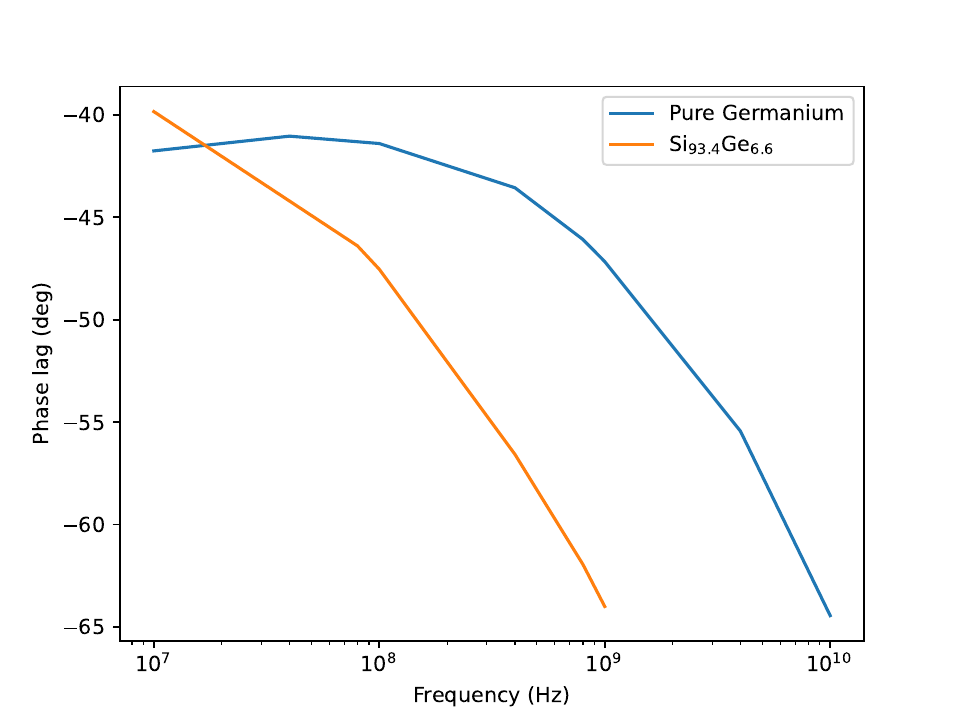}}
\caption{(a) Temperature phase lag in 1D-FDTR in germanium at $T$ = 50 K. The RTA-BTE solutions are for different initial phonon distributions. (b) Temperature phase lag for Si$_{93.4}$Ge$_{6.6}$ in 1D-FDTR obtained from RTA-BTE solutions at $T$ = 300 K.}\label{fig:4}
\end{figure}

A sizeable discrepancy between the phase lag of temperature obtained from solutions to RTA-BTE (green line in Figure~\ref{fig:4}a) and HHE (blue line in Figure~\ref{fig:4}a) is found in germanium at $T$ = 50 K. This discrepancy, in part, is due to the failure of the HHE at high heating frequencies. As shown in Figure~\ref{fig:2}, the BTE solutions begin to deviate from the diffusion and HHE solutions at low to intermediate frequencies ($\omega \tau \approx 0.01$). This failure is understood to arise from the breakdown of the assumptions in deriving the HHE which are only valid for very small space and time Knudsen numbers. Consequently, the HHE will not be valid for $ \omega \tau>1$, which is the regime where temperature waves, if any, are to be found.

We also investigate the impact of the initial phonon distribution on phase lag. Similar to what was observed in ~\cite{chiloyan2020thermal}, by initially exciting only optical modes or modes with a mean free paths greater than 1 $\mu$m, we observe significant changes in phase lag. This suggests that the RTA-BTE contains ample flexibility in modeling deviations from diffusion in germanium. 

More interestingly, the RTA-BTE qualitatively recovers the non-monotonic behaviour of the phase lag in both pure crystals and alloys (Figure~\ref{fig:4}b), indicating that the use of phase lag alone as the experimental observable is not sufficient for the detection of second sound. 

\begin{figure}
\captionsetup[subfloat]{farskip=1pt,captionskip=1pt}
\centering     
\subfloat[]{\label{fig:5a}\includegraphics[width=0.3\textwidth]{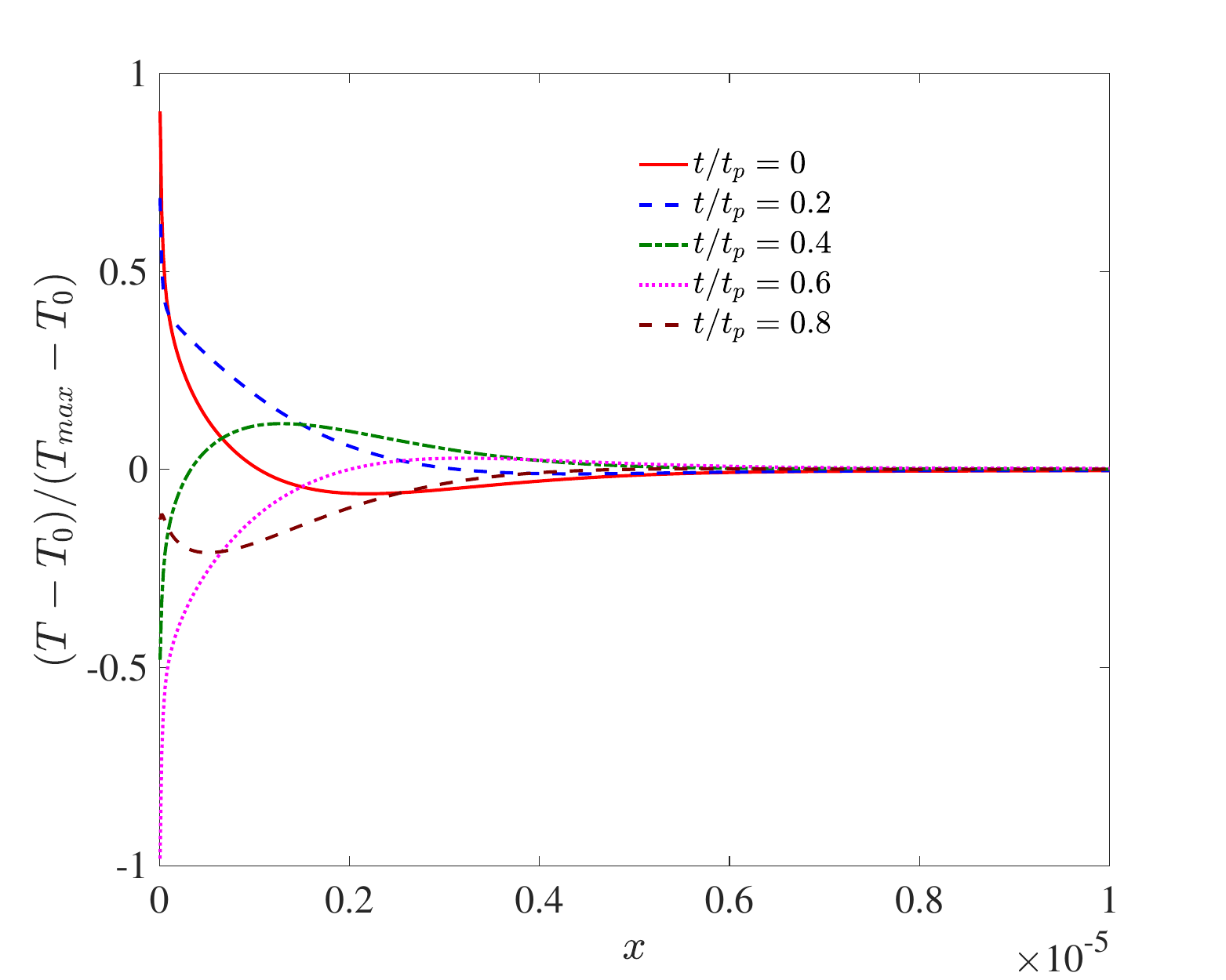}}
\subfloat[]{\label{fig:5b}\includegraphics[width=0.3\textwidth]{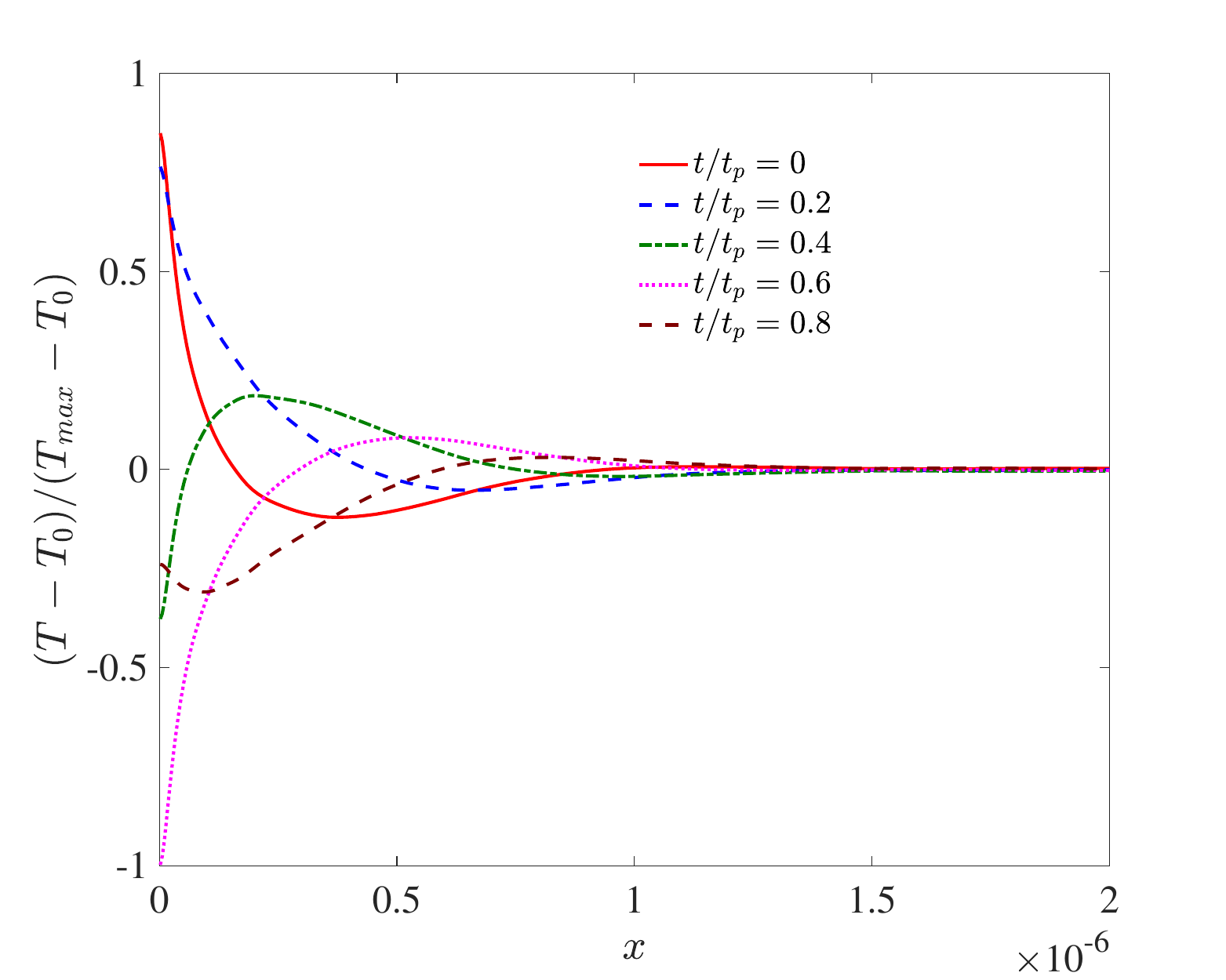}}
\subfloat[]{\label{fig:5c}\includegraphics[width=0.3\textwidth]{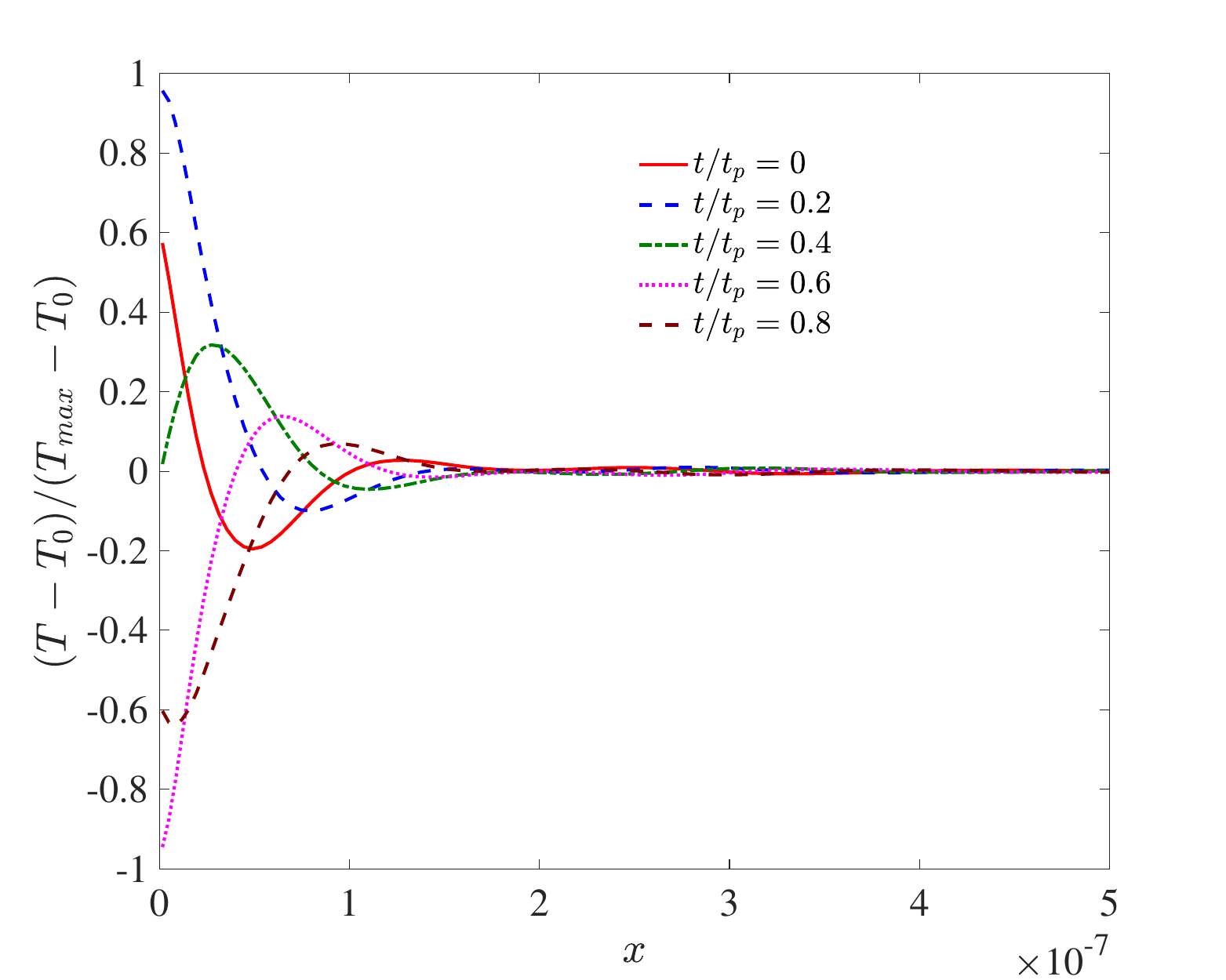}}

\caption{1D-FDTR RTA-BTE solutions for germanium at $T$ = 50 K for heating frequencies of (a) 100 MHz (b) 1 GHz (c) 10 GHz.}\label{fig:5}
\end{figure}

Finally, we report the temperature RTA-BTE profiles (Figure~\ref{fig:5}) at $T$ = 50 K. At $f$ = 100 MHz and 1 GHz, we see that the deviation of phase lag from the diffusion prediction does not necessarily correspond to a wave-like temperature response. It is only  it is only at very high frequencies, $f$ = 10 GHz, which is very near or in the ballistic regime (and consequently far from the validity of the HHE), that wave-like features in the temperature profile can be observed.

\section{Discussion}

Given these observations, how are we to explain the experimental data in Beardo et al.? First, it should be emphasized that temperature waves have not been directly observed, but rather a proxy metric (i.e., phase lag) is interpreted within a model that assumes \textit{a priori} wave-like behavior that is limited to $\omega \tau<0.1$. Past FDTR and time domain thermoreflectance (TDTR) experiments have been modeled within the BTE ~\cite{regner2014analytical,koh2014nonlocal,yang2015heating,hua2017experimental} and the solutions to the BTE obtained here contain the flexibility to predict (or fit) the phase lag.


Second, the `spectrality' of the heat source is not captured in the current HHE. In previous analyses of FDTR experiments, it was shown that the inclusion of a two temperature model for the metal transducer removed much of the non-diffusive behaviour of the substrate~\cite{wilson2014anisotropic,regner2015interpretation}. Notably, by modeling of the transducer, the frequency domain representation of TDTR measurements were captured by a diffusive model up to a modulation frequency of 1 GHz~\cite{collins2014examining}. While there is no metal transducer in Ref.~\cite{beardo2021observation}, the thermalization of hot electrons requires careful treatment~\cite{tong2021toward}. The computational cost of numerically simulating the 3D experimental configuration with the inclusion of electron dynamics remains impractical, nonetheless, it has not been ruled out that the addition of such physics \textit{without invoking second sound or phonon hydrodynamics} in the DUGKS-based solutions to the RTA-BTE would recover the experimental data.

Finally, and most crucially, our BTE solutions demonstrate the absence of a microscopic mechanism for the occurrence of measurable second sound in germanium: the lack of strong normal scattering eliminates the occurrence of drifting second sound, the eigenvalues of the collision matrix eliminate the possibility of driftless second sound and the lack of resonance in temperature response in the mid to high heating frequencies eliminates the `other type' of second sound. 

These issues with the experimental interpretation bring forth the necessity of a `smoking gun' for the observation non-drifting second sound. A second sound `smoking gun' at a minimum would require an analysis that demonstrates the failure of the diffusion, ballistic and RTA-BTE solutions to describe the experimental observable. Current experimental observables are limited to measurements of macroscopic quantities (i.e., temperature). For second sound and phonon hydrodynamics, a collective behaviour in the phonon distribution functions must occur. While current experiments do not probe the phonon distribution functions, there exist techniques that could be modified to track such quantities. Recently, ultrafast electron diffuse scattering experiments have demonstrated the capability to resolve the phonon distributions functions over short timescales~\cite{de2019time}. Such experiments would provide insight into the validity of the LBTE to describe thermal transport at short time and space scales.

\section{Conclusion}

By analyzing the solutions to the LBTE and RTA-BTE, we have shown that the microscopic conditions for drifting, driftless and `other types' of second sound are not satisfied in germanium for $T$ = 50 K to 300 K. However, the story does not end here; quoting Hardy~\cite{hardy1970phonon}: `If none of the sets of conditions for the existence of second sound discussed here are satisfied for a particular material and temperature range, it does not follow that the applicability of the diffusion equation for heat extends to arbitrarily rapidly varying processes. It means only that the range of applicability of the diffusion equation cannot necessarily be extended by simply adding on a term which changes it to a damped wave equation. Nothing in the present discussion excludes the possibility of there being even more types of second sounds than the three suggested here.' Further experimental and theoretical work is needed to bring closure to this puzzle.

\section{Acknowledgments}

We acknowledge discussions with Alexei Maznev. This work was supported by the NSERC Discovery Grants Program under Grant No. RGPIN-2021-02957 (S.H.) and by the National Natural Science Foundation of China (12147122) and the China Postdoctoral Science Foundation (2021M701565) (C.Z.).

\newpage
\bibliographystyle{apsrev}
\bibliography{refs.bib}

\newpage
\section{Supplemental Material: On the question of second sound in germanium: A theoretical viewpoint}

\newpage
\section{DFT calculations}
The DFT calculation parameters used in this work are the following: for the DFPT portion, a 16 $\times $ 16 $\times$ 16 Monkhorst-Pack $k$ mesh with a kinetic energy cutoff of 50 Ry and a convergence criteria of 1.0E-12 Ry is used. For the supercell calculations, a 4 $\times $ 4 $\times$ 4 supercell was used such that third order force constants up to the fifth nearest neighbor could be obtained and only wavefunctions at the gamma point were calculated. Both Ge.pz-bhs.UPF and Ge.pz-n-nc.UPF pseudopotentials were tested yielding a negligible difference between thermal conductivity estimates. The DFPT calculations were done with a 6 $\times $ 6 $\times$ 6 $q$ mesh. Interpolation was done on a 48 $\times $ 48 $\times$ 48 $q$ mesh with a Gaussian smearing parameter of 0.1 for the Kronecker delta approximation to yield convergence of the thermal conductivity. All DFT calculations for germanium were done with the quantum-ESPRESSO package~\cite{giannozzi2009quantum}. The germanium properties have been experimentally verified in Ref.~\cite{dennett2018thermal}. For graphite, the details of the \textit{ab initio} calculations can be found in Ref.~\cite{ding2018phonon}. The macroscopic relaxation time, $\tau$, was calculated using the obtained \textit{ab initio} inputs following the procedure described in Ref.~\cite{beardo2021observation}. Relevant data and source code can be found at https://github.com/schuberm/.

\begin{figure}[H]
\centering     
\subfloat[]{\label{fig:a}\includegraphics[width=0.45\textwidth]{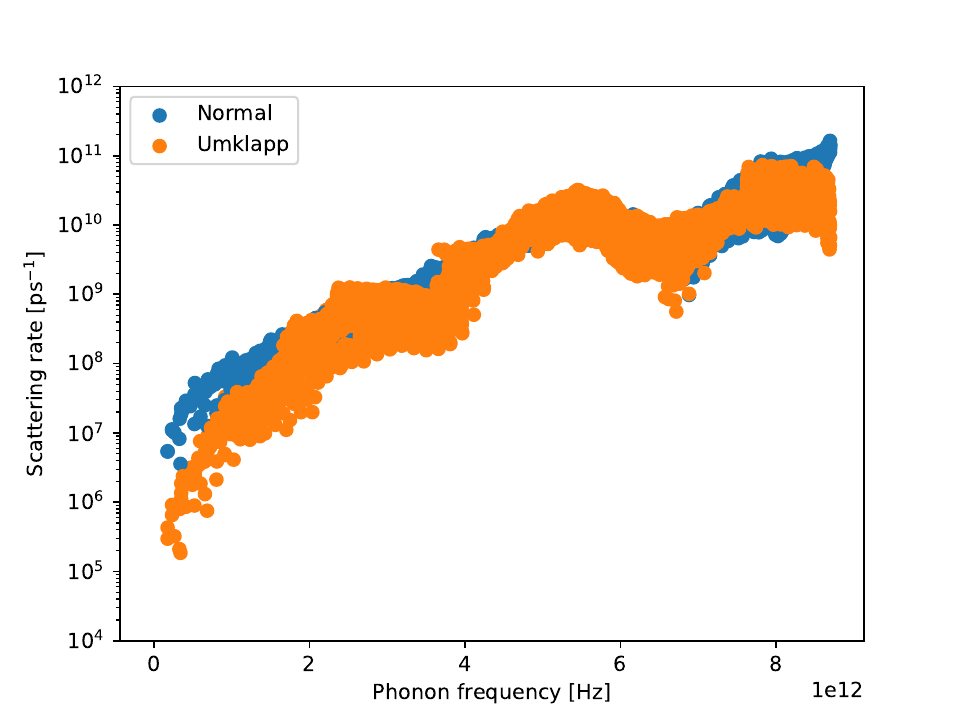}}
\subfloat[]{\label{fig:b}\includegraphics[width=0.45\textwidth]{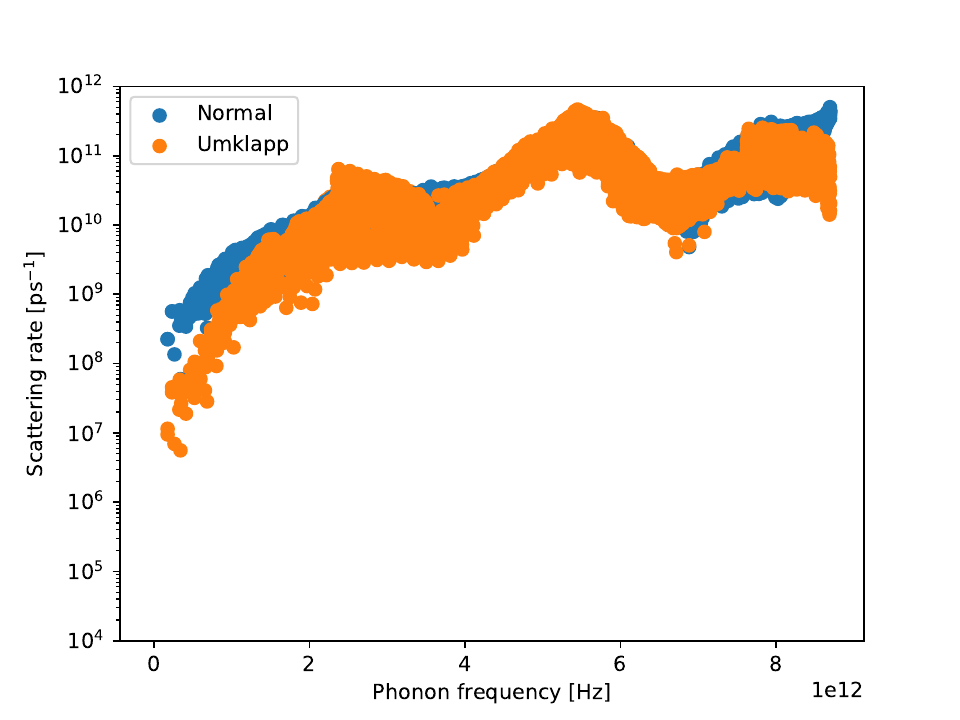}}
\caption{Normal and Umklapp scattering rates in germanium at (a) $T$ = 50 K and (b) 300 K.}
\end{figure}

\begin{figure}[H]
\centering     
\subfloat[]{\label{fig:a}\includegraphics[width=0.9\textwidth]{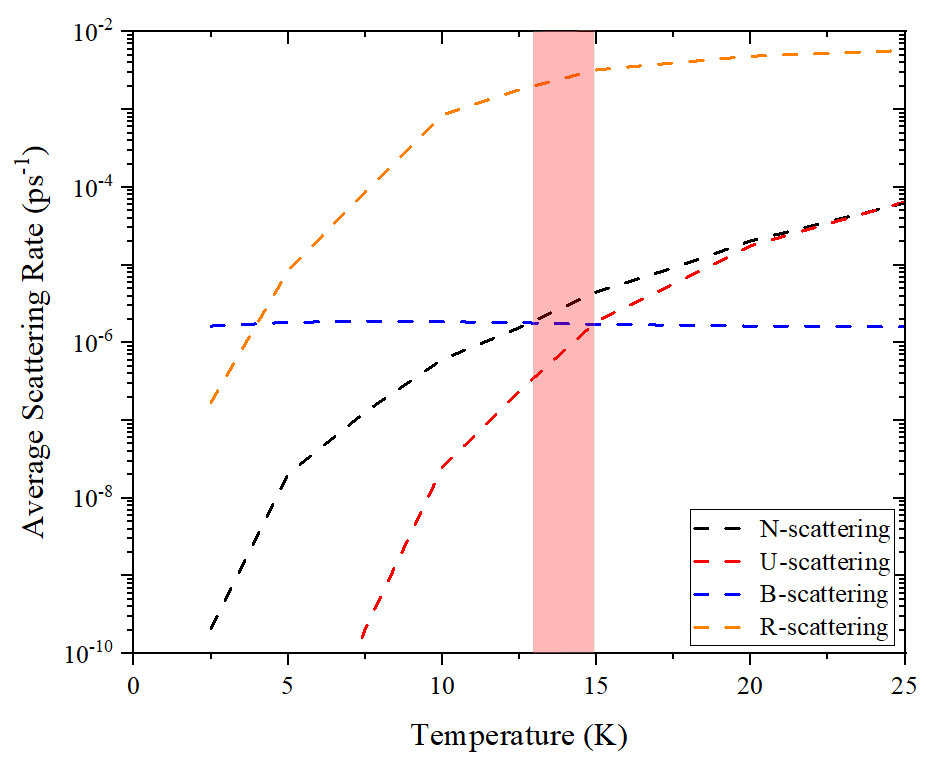}}
\caption{Application Guyer and Krumhansl's criteria~\cite{guyer1966thermal}: $\langle\tau_N^{-1}\rangle > \langle\tau_B^{-1} \rangle > \langle\tau_{U,R}^{-1} \rangle$, where $N,U,R,B$ corresponds to the Normal, Umklapp, resistive (i.e., Umklapp and isotope) and boundary scattering rates respectively and $\langle \rangle$ denotes thermal averages, to estimate the temperature and length window for second sound in germanium. For pure germanium, a window (shaded pink region) emerges at $T$ $<$ 15 K for lengths greater than 1 mm. However, once isotope scattering is included, the hydrodynamic window disappears.}
\end{figure}

\newpage{}

\section{Discrete unified gas kinetic scheme}

Assuming the temperature rise due to the heating is small compared to the background temperature $T_{\text{ref}}$ such that a local temperature $T$ exists, and the phonon Boltzmann transport equation (BTE) under the single-mode relaxation time approximation (RTA) could be written as~\cite{ChenG05Oxford}
\begin{align}
\frac{\partial g_n}{\partial t} + \vec{ \bm{v}_n} \cdot \nabla  g_n = \frac{ g_n^0 -g_n}{ \tau_n } +  Q_n,
\label{eq:pBTE}
\end{align}
where $g_n$ is the deviational phonon energy density for mode $n$, $\vec{ \bm{v}_n}$ is the phonon group velocity, $\tau_n=\tau_n (T_{\text{ref}})$ is the effective relaxation time, $g^0_n =C_n (T- T_{\text{ref}})$ is the deviational equilibrium phonon energy density and $C_n$ is the mode-dependent specific heat.
$Q_n$ is the mode-dependent heat source, which depends on the spatial position $\bm{x}$ and time $t$.
The local energy density $U$, temperature $T$ and heat flux $\bm{q}$ could be obtained by taking the moment of phonon distribution functions,
\begin{align}
U &= \int g_n  d\bm{K},  \label{eq:energy} \\
T &= \frac{\int g_n  d\bm{K}}{\int C_n d\bm{K} } + T_{\text{ref}},  \label{eq:equivalenttemperature}  \\
\bm{q} &= \int \vec{ \bm{v}_n} g_n  d\bm{K},  \label{eq:heatflux}
\end{align}
where $d\bm{K}$ is an integral over the whole wave vector space.
A pseudo-temeprature $T_p$ is introduced to ensure the energy conservation of the phonon scattering term,
\begin{align}
\int \frac{ g_n^0(T_p) -g_n}{ \tau_n (T_{\text{ref}}) } d\bm{K} =0,
\end{align}
so that
\begin{align}
T_{p} = \frac{\int g_n/ \tau_n  d\bm{K}}{\int C_n / \tau_n d\bm{K}} + T_{\text{ref}}.
\end{align}

The discrete unified gas kinetic scheme (DUGKS)~\cite{guo_progress_DUGKS,GuoZl16DUGKS} is introduced to solve Eq.~\eqref{eq:pBTE} with $ab~initio$ input.
The phonon dispersion and scattering in the whole first Brillouin zone are used, and no isotropic wave vector space is assumed, which is different from previous studies~\cite{GuoZl16DUGKS,LUO2017970,zhang_discrete_2019}.
The finite volume method is used to discretize the spatial space, the trapezoidal quadrature is used for the time integration of the phonon scattering term and heat source term, while the mid-point rule is used for the flux term.
Then Eq.~\eqref{eq:pBTE} in integral form over a control volume can be written as follows,
\begin{equation}
g_{i,n}^{m+1}-g_{i,n}^{m} + \frac{\Delta t}{V_i} \sum_{j \in N(i)} \left( \vec{ \bm{v}_n} \cdot \mathbf{n}_{ij} g_{ij,n}^{m+1/2} S_{ij} \right)  =\frac{\Delta t}{2}  \left( \frac{g_{i,n}^{0,m+1} -g_{i,n}^{m+1} }{\tau_{n}^{m+1} } +Q_{i,n}^{m+1}  + \frac{g_{i,n}^{0,m} -g_{i,n}^{m} }{\tau_{n}^{m} }+Q_{i,n}^{m} \right),
\label{eq:dpBTE}
\end{equation}
where $V_i$ is the volume of the cell $i$, $N(i)$ denotes the sets of neighbor cells of cell $i$, $ij$ denotes the interface between cell $i$ and cell $j$, $S_{ij}$ is the area of the  interface $ij$, $\mathbf{n}_{ij}$ is the normal unit vector of the interface $ij$ directing from cell $i$ to cell $j$; $\Delta t$ is the time step from time $t_m$ to $t_{m+1}=t_{m}+ \Delta t$, where $m$ is an index of the time steps.
In order to remove the implicitness, two new distribution functions are introduced and defined as
\begin{equation}
\tilde{g}=g-\frac{\Delta t}{2}( \frac{g^0-g}{\tau}  +Q ),
\label{eq:I1}
\end{equation}
\begin{equation}
\tilde{g}^{+}=g + \frac{\Delta t}{2}( \frac{g^0-g}{\tau}  +Q ).
\label{eq:I2}
\end{equation}
Then Eq.~\eqref{eq:dpBTE} can be expressed as
\begin{equation}
\tilde{g}_{i,n}^{m+1}-\tilde{g}_{i,n}^{+,m}+ \frac{\Delta {t}}{V_i} \sum_{j \in N(i)} \left( \vec{ \bm{v}_n}  \cdot \mathbf{n}_{ij} g_{ij,n }^{m+1/2} S_{ij} \right)  =0.
\label{eq:dpBTE2}
\end{equation}
From Eq.~\eqref{eq:dpBTE2}, it can be found that the phonon distribution functions at the cell interface at the mid-point time step $g^{m+1/2}_{ij,n}$ and at the cell center at the next time step $g^{m+1}_{i,n}$ are both needed to be calculated in the DUGKS.

Firstly, we need to obtain the phonon distribution functions at the cell interface $g^{m+1/2}_{ij,n}$.
Different from direct numerical interpolation used in the discrete ordinate method, in the DUGKS, the phonon BTE is employed on the reconstruction of the distribution function at the cell interface.
This is achieved by integrating Eq.~\eqref{eq:pBTE} from time $t_m$ to $t_{m+1/2}=t_{m}+ \Delta t/2$ along the characteristic line with the end point $\bm{x}_{ij}$ locating at the center of the cell interface $ij$ between cell $i$ and cell $j$,
\begin{equation}
g_{n}^{m+1/2}(\bm{x}_{ij})-g_{n}^{m}(\bm{x}_{ij}')
=\frac{\Delta t}{4} \left[  \left.\   \left( \frac{g_{n}^{0,m+1/2} -g_{n}^{m+1/2} }{\tau_{n}^{m+1/2} } +Q_n^{m+1/2} \right)  \right|_{\bm{x}=\bm{x}_{ij}}    +  \left.\  \left(  \frac{g_{n}^{0,m } -g_{n}^{m } }{\tau_{n}^{m } } +Q_n^{m} \right) \right|_{\bm{x}=\bm{x}_{ij}'}  \right],
\label{eq:fBTE}
\end{equation}
where $\bm{x}_{ij}'= \bm{x}_{ij}-\vec{ \bm{v}_n} \Delta t/{2}$.
Equation~\eqref{eq:fBTE} can be reformulated as follows,
\begin{equation}
\bar{g}_{n}^{m+1/2}(\bm{x}_{ij})-\bar{g}_{n}^{+,m}(\bm{x}_{ij}') =0,
\label{eq:fBTE2}
\end{equation}
where
\begin{equation}
\bar{g}=g-\frac{\Delta t}{4}( \frac{g^0-g}{\tau}  +Q ),
\label{eq:I3}
\end{equation}
\begin{equation}
\bar{g}^{+}=g+\frac{\Delta t}{4}( \frac{g^0-g}{\tau}  +Q ).
\label{eq:I4}
\end{equation}
$\bar{g}_{n}^{+,m}(\bm{x}_{ij}')$ in Eq.~\eqref{eq:fBTE2} is reconstructed by numerical interpolation,
\begin{equation}
\bar{g}_{n}^{+,m}(\bm{x}_{ij}') = \bar{g}_{n}^{+,m}(\bm{x}_{c}) + (\bm{x}_{ij}'-\bm{x}_{c}) \bm{\sigma}_{c},
\label{eq:slope}
\end{equation}
where $\bm{\sigma}_{c}$ is the spatial gradient of the distribution function $\bar{g}_{n}^{+,m}(\bm{x}_{c})$  in the cell $c$.
If $\vec{ \bm{v}_n}   \cdot \bm{n}_{ij} >0$, $c=i$; else $c=j$.
The van Leer limiter is adopted to determine the gradient to ensure the numerical stability and accuracy.

Combining Eqs.~\eqref{eq:fBTE2},~\eqref{eq:I4} and~\eqref{eq:slope}, the new distribution function $\bar{g}_{n}^{m+1/2}(\bm{x}_{ij})$ at the cell interface at time $t_{m+1/2}$ can be obtained.
Make a transformation of Eq.~\eqref{eq:I3},
\begin{align}
\frac{ \bar{g} + \frac{\Delta t}{4} Q - g^0  }{4 \tau +\Delta t } =\frac{g^0 -g}{ 4 \tau}.
\end{align}
According to the energy conservation principle of the scattering term,
\begin{align}
\int \frac{ \bar{g}_n + \frac{\Delta t}{4} Q_n - g^0_n  }{4 \tau_n +\Delta t } d\bm{K} =\int  \frac{g^0_n -g_n}{ 4 \tau_n} d\bm{K}  =0.
\end{align}
Then
\begin{align}
T_{ij,p}^{m+1/2} =\left(  \int \frac{\bar{g}_{ij,n}^{m+1/2} + \frac{\Delta t}{4} Q_{ij,n}^{m+1/2} }{4 \tau_n +\Delta t }  d\bm{K}  \right) / \left(  \int \frac{C_n}{4 \tau_n +\Delta t }  d\bm{K}   \right).
\end{align}
Once $T_{ij,p}^{m+1/2}$ is known, $g^{0,m+1/2}_{n,ij}$ can be obtained, and then the original phonon distribution function at the cell interface $g^{m+1/2}_{n,ij}$ can be calculated by Eq.~\eqref{eq:I3}.

Similar treatment can be conducted when updating the phonon distribution functions at the cell center at the next time step $g^{m+1}_{i,n}$.
Details are shown below.
Once $g^{m+1/2}_{ij,n}$ and $g^{0,m+1/2}_{ij,n}$ are known, $\tilde{g}_{i,n}^{m+1}$ can be updated by Eq.~\eqref{eq:dpBTE2}.
Make a transformation of Eq.~\eqref{eq:I1},
\begin{align}
\frac{ \bar{g} + \frac{\Delta t}{2} Q - g^0  }{2 \tau +\Delta t } =\frac{g^0 -g}{ 2 \tau}.
\end{align}
According to the energy conservation principle of the phonon scattering term,
\begin{align}
\int \frac{ \tilde{g}_n + \frac{\Delta t}{2} Q_n - g^0_n  }{2 \tau_n +\Delta t } d\bm{K} =\int  \frac{g^0_n -g_n}{2 \tau_n} d\bm{K}  =0.
\end{align}
Then the pseudo-temperature at the cell center $T_{i,p}^{m+1}$ can be updated,
\begin{align}
T_{i,p}^{m+1} =\left(  \int \frac{\tilde{g}_{i,n}^{m+1} + \frac{\Delta t}{2} Q_{i,n}^{m+1}}{2 \tau_n +\Delta t }  d\bm{K}  \right) / \left(  \int \frac{C_n}{2 \tau_n +\Delta t }  d\bm{K}   \right).
\label{eq:centerTp}
\end{align}
Once $T_{i,p}^{m+1}$ is known, $g^{0,m+1}_{i,n}$ can be calculated, and finally the original phonon distribution function at the cell center at the next time step $g^{m+1}_{n,i}$ can be calculated based on Eq.~\eqref{eq:I1}, and the associated macroscopic variables at the next time step could be updated based on Eqs.~(\ref{eq:energy},\ref{eq:equivalenttemperature},\ref{eq:heatflux}), too.
Above procedures are the main evolution processes of the phonon distribution function in the DUGKS.
The performance of the DUGKS has been validated in previous studies~\cite{guo_progress_DUGKS,GuoZl16DUGKS,LUO2017970,zhang_discrete_2019}.

When the mode-dependent BTE with $ab~initio$ calculations input is simulated, the heat source is mode-dependent
\begin{align}
Q_n =\frac{C_n}{ \int C_n d\bm{K} } \dot{S}(\bm{x},t),
\end{align}
where $\dot{S}$ is the heat source at the macroscopic level.
When the linear phonon dispersion and gray model is used, the specific heat, group velocity, heat source and relaxation time are all frequency-independent.
For three-dimensional materials, we have $\vec{v}=|\vec{v}| \bm{s}$, where $\bm{s}=\left( \cos \theta, \sin \theta \cos \varphi, \sin \theta  \sin \varphi  \right)$ is the unit directional vector, $\theta \in [0,\pi]$, $\varphi \in [0,2\pi]$ is the azimuthal angle.
The $\cos \theta \in [-1,1]$ is discretized with the $N_{\theta}$-point Gauss-Legendre quadrature, while the azimuthal angular space $\varphi \in [0,\pi]$ (due to symmetry) is discretized with the $\frac{N_{\varphi}}{2}$-point Gauss-Legendre quadrature.

For all quasi-1D periodic steady problems in this paper, the simulated system size is large enough to ensure the semi-infinite boundary.
Non-uniform cells are used to discrete the spatial domain, and the cell size increases form the left to the right with a fixed ratio $r_{size}$.
When $\epsilon < 10 ^{-6}$, the thermal system reaches periodic steady state, where
\begin{align}
\epsilon =  \frac{ \sum_{i=1}^{N_{cells}} \left( (T(x_i,t + t_p) -T(x_i,t) \right)^2   }{\sum_{i=1}^{N_{cells}} (\Delta T)^2 },
\end{align}
where $N_{cells}$ is the total discretized spatial cells, $t_p$ is the heating period, and $x_i$ is the spatial position of the cell $i$.

\section{Quasi-1D frequency domain thermoreflectance}

For quasi-1D frequency domain thermoreflectance (FDTR) geometry~\cite{beardo_observation_2021}, we have
\begin{align}
\dot{S}(\bm{x},t)= A \exp(-x/l_{pump})\cos( 2 \pi t/ t_p ),
\end{align}
where $A$ is the amplitude of the heat source.
The left boundary is the diffusely reflecting adiabatic boundary~\cite{GuoZl16DUGKS} and the right is assumed semi-infinite with temperature $T_{\text{ref}}$.

\begin{figure}[htb]
 \centering
 \includegraphics[scale=0.4,clip=true]{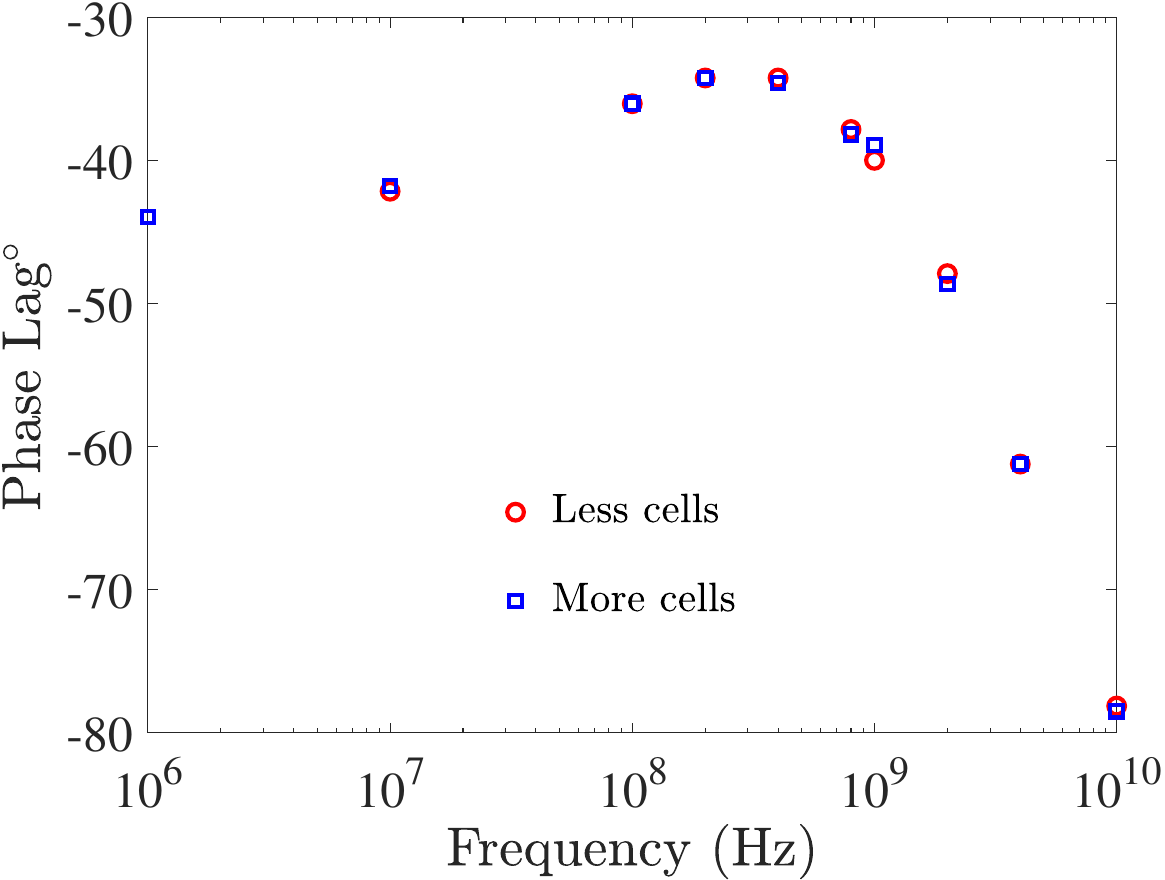}
 \caption{The heating frequency dependent phase lag in quasi-1D FDTR~\cite{beardo_observation_2021}. The mesh-independence is tested with different discretized cells. The minimal cell size of ``Less cells'' is 4 times larger than that of ``More cells''. }
 \label{duibiphaseLag}
\end{figure}

The linear phonon dispersion and gray model is used, and the parameters of Germanium at $300$ K in Ref.~\cite{beardo_observation_2021} are adopted, where $C= 1.60 \times 10^6 $ J/(m$^3 \cdot$K), $|\vec{v}|= 440$ m/s, $\tau=500$ ps, $l_{pump}=15$ nm.
We set $N_{\theta} \times N_{\varphi} =100 \times 8$, $r_{size}=1.01$-$1.03$.
The transient heat conduction with different heating frequency $f=1/t_p$ is simulated and the results are shown in~\ref{duibiphaseLag}.
It can be found that the phase lag between the temperature and heat source changes non-monotonously with the heating frequency.
\begin{figure}[htb]
 \centering
 \includegraphics[scale=0.4,clip=true]{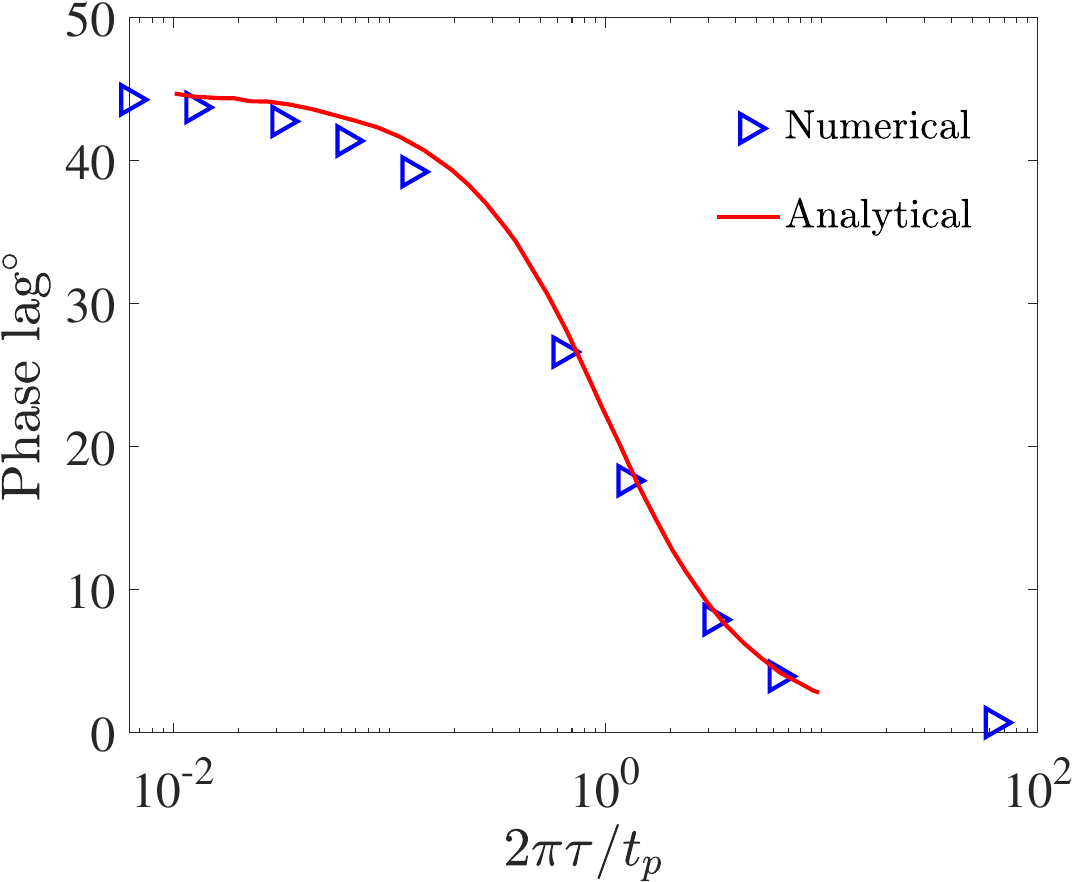}
 \caption{The heating frequency dependent phase lag predicted by the numerical and analytical solutions~\cite{PhysRevB.91.165311}.}
 \label{phaselag1Dgray}
\end{figure}

\section{Quasi-1D transient heat conduction with time varying boundary}

The temperature at the left boundary is fixed and changed with time~\cite{PhysRevB.91.165311}, i.e.,
\begin{align}
T(x=0,t)= T_{\text{ref}} +\Delta T \cos( 2 \pi t/ t_p ),
\end{align}
where $t_p$ is the heating period, $\Delta T$ is the temperature increment.
The right boundary is assumed semi-infinite with temperature $T_{\text{ref}}$.
The isothermal boundary conditions are used for both two boundaries~\cite{GuoZl16DUGKS}.

When the linear phonon dispersion and gray model is used, the ratio between the relaxation time and heating period $\tau/t_p$ totally control the transient heat conduction based on the dimensional analysis of the RTA-BTE~\eqref{eq:pBTE}.
We set $N_{\theta} \times N_{\varphi} =100 \times 8$, $r_{size}=1.04$.
The numerical results predicted by the DUGKS with different $\tau /t_p$ are shown in~\ref{phaselag1Dgray}.
It can be found that in this problem, the phase lag between the temperature and heat flux decreases monotonously with $\tau / t_p$.

\newpage
\section{Appendix A}

\begin{figure}[H]
\centering  
\includegraphics[width=0.75\textwidth]{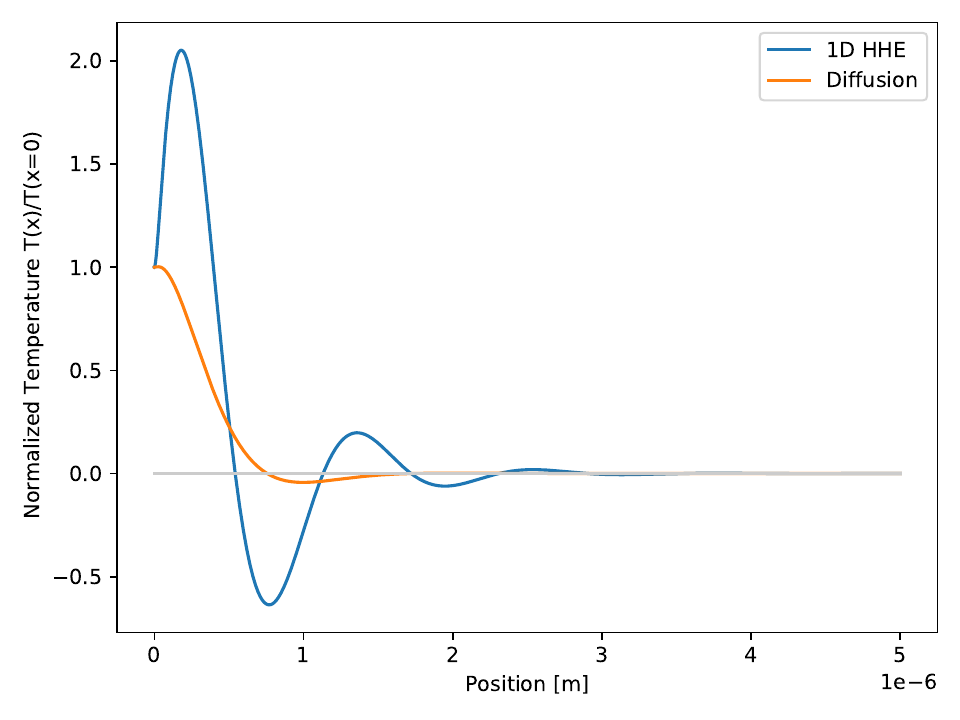}
\caption{Example temperature profile for the 1D-FDTR geometry for the diffusion and HHE.}
\end{figure}\label{}

\newpage
\section{Appendix B}

\begin{figure}[H]
\centering  
\includegraphics[width=0.75\textwidth]{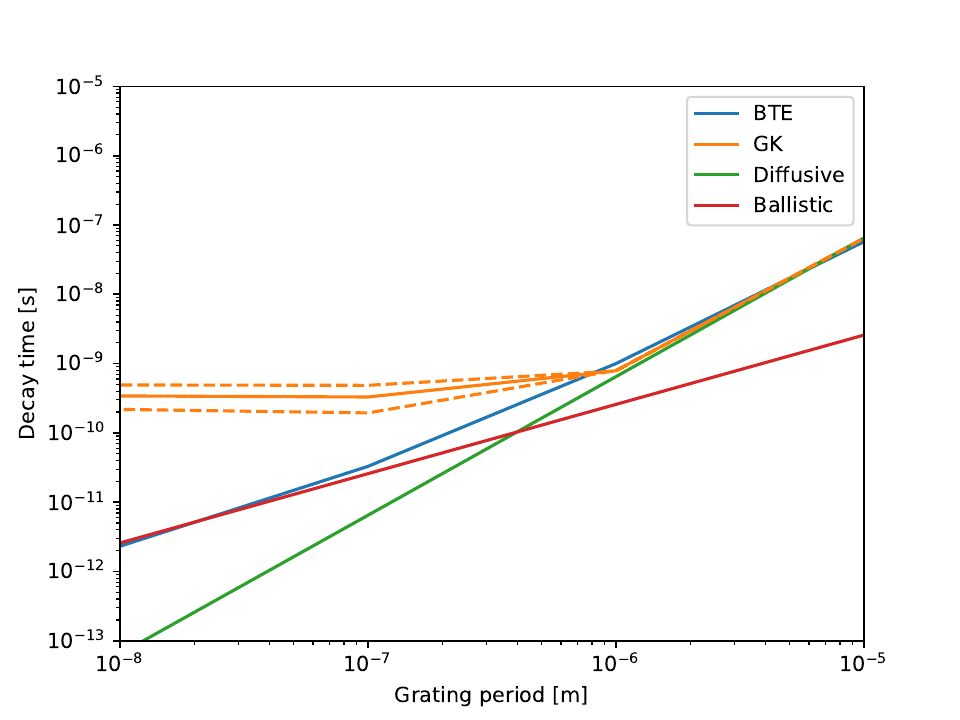}
\caption{Decay times for germanium at T = 300 K in the 1D TTG for the diffusion, ballistic, LBTE and the Gruyer-Krumhansl (GK) equations. Dashed lines are $\pm$ $\%$ 20 of the non-local length $l= 100$ nm.}
\end{figure}\label{}

\newpage
\section{Appendix C}

\begin{figure}[H]
\centering  
\includegraphics[width=0.75\textwidth]{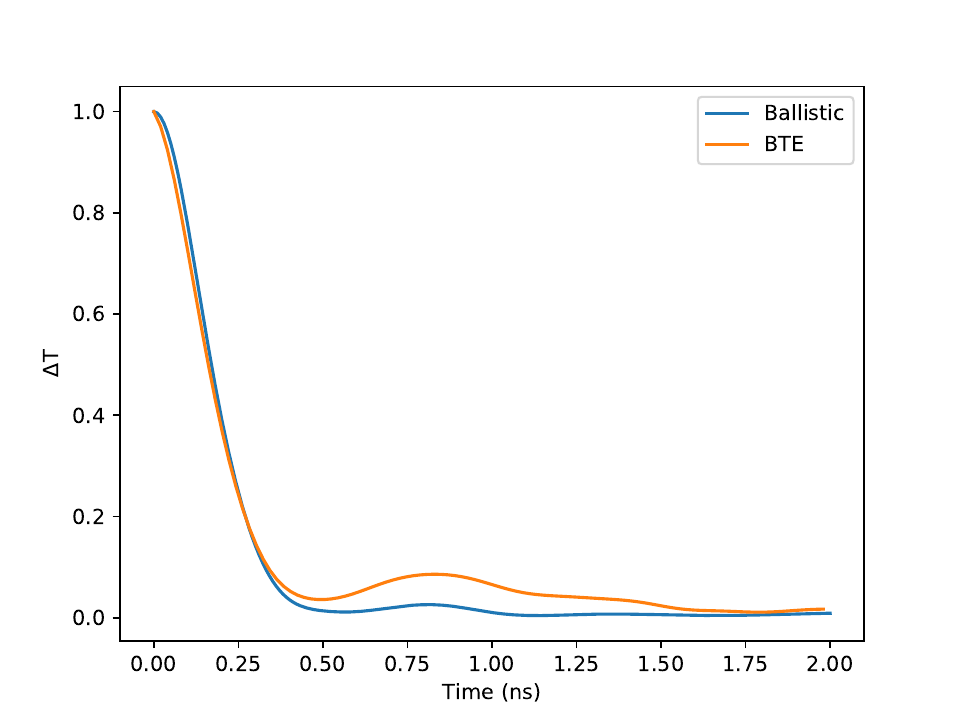}
\caption{Temperature response for the 1D-TTG in germanium at T = 50 K for L = 1 $\mu$m.}
\end{figure}\label{}

\newpage
\section{Appendix D}

\begin{figure}[H]
\centering  
\includegraphics[width=0.75\textwidth]{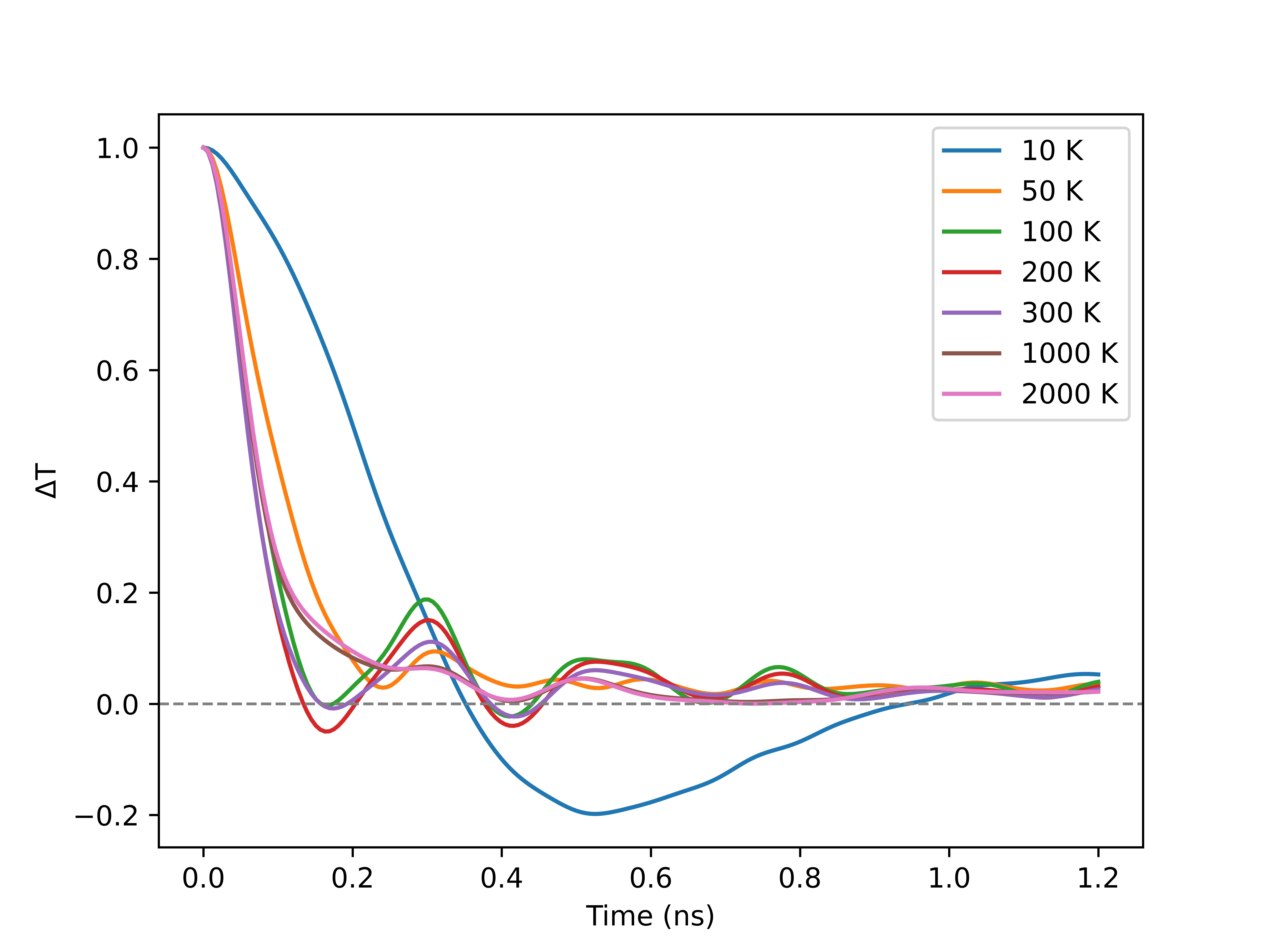}
\caption{Ballistic temperature responses in graphite for 1D-TTG. These oscillations are attributed to the first speeds of sound.}
\end{figure}\label{}

\end{document}